\documentclass[10pt,journal,compsoc]{IEEEtran}

\ifCLASSOPTIONcompsoc
  \usepackage[nocompress]{cite}
\else
  \usepackage{cite}
\fi

\usepackage{tabularx}
\usepackage{booktabs}
\usepackage{makecell}
\usepackage{graphicx}
\usepackage{color, colortbl}
\usepackage{longtable}

\newcolumntype{L}{>{\raggedright\arraybackslash}X}
\definecolor{Gray}{gray}{0.9}

\begin{document}

\title{Release early, release often,\\ \textit{and watch your users' emotions}}

\author{Daniel~Martens 
        and~Walid~Maalej}

\markboth{IEEE Theme Issue: Sentiment and Emotion in Software Engineering}%
{IEEE Theme Issue: Sentiment and Emotion in Software Engineering}

\IEEEtitleabstractindextext{%
\begin{abstract}
App stores are highly competitive markets, sometimes offering dozens of apps for a single use case. Unexpected app changes such as a feature removal might incite even loyal users to explore alternative apps. Sentiment analysis tools can help monitor users' emotions expressed, e.g., in app reviews or tweets. We found that these emotions include four recurring patterns corresponding to the app releases. Based on these patterns and online reports about popular apps, we derived five release lessons to assist app vendors maintain positive emotions and gain competitive advantages. 
\end{abstract}

\begin{IEEEkeywords}
Sentiment analysis, app reviews, user emotions, emotional patterns, release lessons
\end{IEEEkeywords}
}

\maketitle

\IEEEdisplaynontitleabstractindextext

\IEEEpeerreviewmaketitle

\IEEEraisesectionheading{\section{Extract Users' Emotions from Feedback}}

\IEEEPARstart{I}{nitially} developed for marketing and political opinion mining, sentiment analysis became popular in many domains including software engineering. Mining emotional sentiment has been used, e.g., to guide developer discussions~\cite{Calefato:2018:SPD:3231288.3231327} or summarize users opinion on app features~\cite{6912257}.

Sentiment analysis tools use natural language processing to extract emotions from text messages. Simple lexicon-based approaches match each text token to dictionaries containing negative or positive words. The dictionaries define  sentiment scores for specific tokens, such as \textit{``I hate[-4] that u need wifi but overall the app is great[+3]''}. Depending on the tool, the scores are combined, e.g., into a single value reflecting the overall emotion expressed.

Table~\ref{tab:sentimenttools} describes state-of-the-art sentiment analysis tools including their approaches, scores, languages supported, technology used, and licenses. For this  investigation we selected SentiStrength~\cite{Thelwall:2010:SSD:1890706.1890713}, which is a common baseline for emotion classification used in many studies (e.g., \cite{Calefato:2018:SPD:3231288.3231327}). 
Its results can  be improved with domain-specific lexicons, where specific terms correlate with other emotions than in general as the term `bug' in software engineering. SentiStrengthSE uses a manually adjusted version of the SentiStrength lexicon. Senti4SD is trained on manually labelled questions, answers, and comments from StackOverflow. Hence, the training data includes more software engineering specific terms.
Most sentiment analysis tools only support the English language, while apps are offered and allow user to  submit feedback in more than 40 languages. 
To analyze these languages further dictionaries need to be created.

\begin{table*}[]
\scriptsize
\renewcommand{\arraystretch}{1.3}
\caption{Overview of Sentiment Analysis Tools}
\label{tab:sentimenttools}
\centering

\begin{tabularx}{\textwidth}{lLLLLL}
\toprule
\textbf{Tool}                        & \textbf{Approach}                                                                         & \textbf{Scores}                                                                                              & \textbf{Languages}                                                                            & \textbf{Technology}          & \textbf{License}                                                                             \\ \rowcolor{Gray}
\midrule
SentiStrength               & Lexicon-based                                                                    & Estimates strength of positive and negative sentiment                                                 & English (Partly tested: Dutch, Finnish, German, Italian, Russian, Spanish, Turkish) & Command-line (Java) & Commercial, closed source (Free for academic research) \\
SentiStrengthSE             & Lexicon-based, ad-hoc heuristics                                                 & Estimates strength of positive and negative sentiment                                                 & English                                                                             & Command-line (Java) & \textit{-- inherits from SentiStrength -- }                                                       \\ \rowcolor{Gray}
Senti4SD                    & Lexicon-based, keyword-based, semantic features (machine learning, word vectors) & Estimates strength of positive and negative sentiment                                                 & English                                                                             & Command-line (R)    & Open source (MIT license)                                                       \\
Vader NLTK                  & Lexicon-based                                                                    & Estimates strength of positive, neutral, negative, and compound sentiment                             & English                                                                             & Python package      & Open source (Apache license)                                                    \\ \rowcolor{Gray}
\makecell[tl]{Google Natural\\ Language API} & Machine-learning (not further specified)                                                                 & Classifies emotion into positive, neutral, and negative                                               & English, Chinese, French, German, Italian, Japanese, Korean, Portuguese, Spanish    & Remote API            & Commercial, closed source                                                            \\
\makecell[tl]{IBM Watson\\ Tone Analyzer}   & Machine-learning (not further specified)                                                                & Identifies presence and estimates strength of analytical, anger, confident, fear, and tentative tones & English, French                                                                     & Remote API            & Commercial, closed source                                                                       \\ 
\bottomrule
\end{tabularx}
\end{table*}


\section{Emotional Patterns in App Reviews}

We applied SentiStrength to 7 million app reviews~\cite{7961885}, corresponding to the 25 top free and paid apps from each of the 23 categories of the US Apple App Store (December 2016). As apps can be listed in several categories, we removed duplicates and only considered apps' main category. In total we analyzed 245 apps. 
We found that users' sentiments are most negative within app reviews of the categories ``photo \& video'' (mean: 0.4), ``entertainment'' (0.6), and ``sports'' (0.9). In contrast, we observed the most positive average sentiments within the categories ``reference'' (2.7), ``education'' (2.4), and ``health \& fitness'' (2.3).

Reviews in app stores include a 1-5 stars rating. We found a moderate correlation between the rating and the sentiment (Pearson correlation coefficient 0.57, Spearman rank correlation 0.56). Compared to star ratings which are often skewed~\cite{6636712}, sentiment scores are more fine grained (e.g., on a range from -5 to +5) and can be calibrated to specific information needs, e.g., by fine-tuning the dictionaries. 
Further, sentiment scores enable comparisons across different platforms. 
Particularly, in channels without explicit user ratings, automated sentiment analysis tools can help quickly assessing  users' overall opinion, as in social media channels (e.g., Twitter). These channels  
are increasingly being used by app vendors to gather feedback from users (e.g., @SpotifyCares, see https://twitter.com/spotifycares), due to their popularity and their ability to exchange information in form of screenshots or video recordings.


Analyzing the sentiment for different apps over time reveals four recurring emotional patterns~\cite{6912257}, as shown in Figure~\ref{fig:patterns}. Each pattern can be related to specific issues or changes within the apps. For example, new app features might increase the average sentiment.
The pattern \textit{consistent emotion} is characterized by a stable negative, neutral, or positive sentiment over time. It can be observed within 15\% of the analyzed apps, such as Spotify or Duolingo.
The pattern \textit{inconsistent emotion} is characterized neither by a constant nor by a clear positive or negative trend. This pattern can be observed for 62\% of the analyzed apps, including WhatsApp.
The pattern \textit{steady decrease/increase} is characterized by a constant negative or positive sentiment trend. A constant negative trend can be observed within 10\% of the analyzed apps, such as CNN or Microsoft Outlook. 
A positive trend can only be observed within 3\% of the apps, such as AccuWeather.
The pattern \textit{emotion drop/jump} is characterized by a sudden change of the sentiment from negative to positive or vice versa. A change from negative to positive can be observed within 15\% of the analyzed apps. A change from positive to negative can be observed within 9\% of the apps, such as Google Mail or OverDrive.


\section{Release Lessons}
Regularly watching users' emotions and identifying corresponding patterns is a first step towards understanding an app's health. We additionally present five release lessons that software practitioners can apply to improve users' emotions and prevent general negative feedback which can lead to the fall of apps~\cite{williams2018modeling}.

We derived the lessons by looking at the release history, the content of user reviews, official vendor's presentations, and technical blogs of several apps corresponding to each pattern. For each lesson we observed at least two indications (e.g., two example apps). Some of the lessons are also supported by recent studies. However, since our lessons are not the result of an in-depth empirical study, we refrain from claiming any generalizability or completeness of lessons. The lessons with their actionable recommendations should encourage and inspire practitioners consider users emotion together with the release frequency and complexity when fine-tuning their release processes.


\subsection{Continuously Analyze User Feedback to Identify and React to Bug Reports and Feature Requests}

Software practitioners should analyze user feedback of released app versions and react to frequently mentioned bug reports and feature requests, especially when similar features are already offered by competitors.

\begin{figure*}[]
\centering
\includegraphics[width=\textwidth]{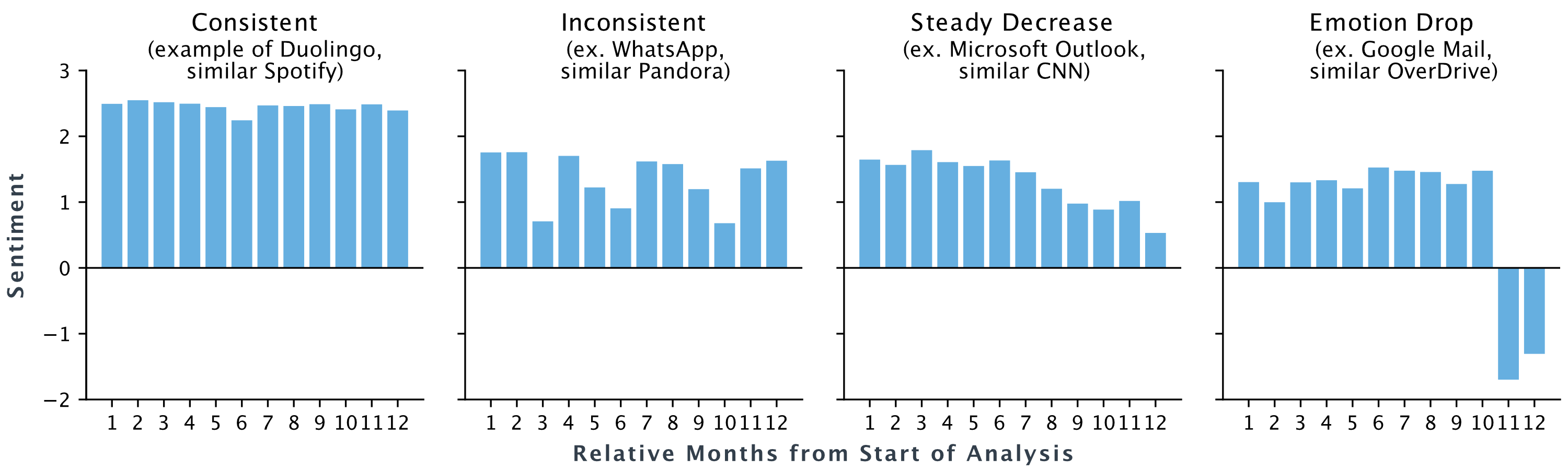}
\caption{Four recurring Emotional Patterns in User Feedback}
\label{fig:patterns}
\end{figure*}

The majority of apps follow the \textit{inconsistent emotion} pattern. These apps are affected by temporary bugs that are quickly fixed by developers. Figure~\ref{fig:patterns} shows that the sentiment of WhatsApp (similar to Pandora) strongly decreases and then restores for single periods of time. In the third month, relative from the start of our analysis, nearly all users report storage issues, e.g., \textit{``Major Storage Issues''}. With the release of an update that fixes the bug after a week, the issue was reported less often. In month 6, users frequently mention crashes after app start. Although these issues only appear for a short period of time, their impact on the overall sentiment is notable since they affect the majority of users who install the update.

Apps that do not react to issues reported by their users are associated with the pattern \textit{steady decrease}. Microsoft Outlook's reviews included diverse emotions until month 6, as shown in Figure~\ref{fig:patterns}. Reviews with positive sentiments such as \textit{``Best email app''} exceeded negative reviews, leading to a stable average sentiment value around 2. The majority of negative reviews is related to issues within the apps, e.g., \textit{``Trash won't empty''} or \textit{``Bug when adding accounts''}. As many competing apps exist, users apparently began to explore alternatives. A user wrote \textit{``In iOS mail, you can copy an attached excel spreadsheets within body of email but outlook doesn't format correctly.''}. Similarly another user reported: \textit{``Although Microsoft has addressed several issue, it's still buggy at times. I just started looking for a replacement app.''}. 

\vspace{10px}
\noindent\fbox{
  \parbox{.95\columnwidth}{
    \textbf{Recommendation \#1:}\vspace{1mm}\\
    Software practitioners should use tools (e.g., https://openreq.eu/) to classify and extract bug reports and feature requests from user feedback. Even when already using automated crash reporting tools, user feedback might include additional non-crashing bugs that cannot be automatically captured. These bugs should be clustered to determine their severity. Bugs that are frequently reported should be quickly fixed by developers before users explore alternative apps. Martin et al.~\cite{7765038} provide a broad overview of the research area app store analysis and existing approaches.
  }
}

\subsection{Frequently Release Small Changes}

If possible, software practitioners should frequently introduce small changes to their apps instead of releasing fundamental changes at once, such as a major redesign of the user interface or the removal of app features. 

For apps introducing major changes at once we observed \textit{emotion drops}. For Google Mail users provided reviews with positive sentiments until the tenth month of our analysis, such as \textit{``Love it more than iPhone mail''}. With an update that applied Android's material design, that iOS users are unfamiliar with, the sentiment suddenly turns negative. Reviews including negative sentiments were often related to usability issues (e.g., \textit{``New update makes you click on each individual email to delete them.''}) or to features removed (e.g., \textit{``Bring back Mark as Unread.''}).
A similar emotion drop can be observed as OverDrive introduced major changes within a single app update. 
The update included multiple bugs, as several users reported \textit{``Can't download books to device''}, \textit{``Buggy, buggy, annoying''}, or \textit{``App crashes with last update''}.

In case of Google Mail, the vendor reacted with weekly updates integrating features requested by users in the reviews succh as \textit{``Select multiple messages [...]''} and \textit{``Mark as read/unread [...]''}. With the release of those updates the sentiment shows a positive trend. For OverDrive,  to restore the sentiment most bugs were  fixed at once with a single app update after a longer period of time.

The frequency of app updates is controversially discussed in the literature. A recent study reports that frequently updated apps receive a significantly lower percentage of negative ratings~\cite{McIlroy2016}. Another study only found a weak correlation, considering negative and positive ratings~\cite{Martin:2016:CIA:2950290.2950320}. However, the study shows that the types of released changes have a varying impact on app ratings. Terms and topics around bug fixes and features occur frequently in the description of impactful releases~\cite{Martin:2016:CIA:2950290.2950320}.

We recommend to consider releasing frequent and small updates to avoid surprising users with unexpected (i.e., too many or major) changes~\cite{McIlroy2016, Martin:2016:CIA:2950290.2950320, 7081842}. Further, we consider frequent releases beneficial, since bugs get fixed faster for apps with shorter release cycles~\cite{6224279}. Also, studies show that app releases lead to an increased amount of ratings and reviews~\cite{6636712, Martin:2016:CIA:2950290.2950320, Martens2019}, allowing  developers to get more feedback and  better understand user needs in highly competitive and dynamic  markets~\cite{williams2018modeling}.

\vspace{10px}
\noindent\fbox{
  \parbox{.95\columnwidth}{
    \textbf{Recommendation \#2:}\vspace{1mm}\\
    High code churn in releases, i.e., the rate at which an app's code evolves, correlates with lower ratings~\cite{7081842} and sentiment scores. Software practitioners should use issue trackers' and version control systems' built-in functionality or external tools to visualize the amount of change introduced (e.g., number of user stories resolved, number of bugs fixed, lines of code implemented). Based on these measures, the severity of changes can be determined to decide whether these should be introduced in separate smaller, more frequent releases. 
  }
}


\subsection{Pre-Release Changes to Subsets of App Users}

Changes should be pre-released to subsets of app users before making these available to everyone. Spotify and Duolingo apply this lesson and are able maintain a consistent positive emotion among their users.

For initial tests, software practitioners should provide access to alpha and beta app versions to voluntary users, as Spotify does (https://bit.ly/2T00cj1). The alpha version is updated almost daily and may be affected by stability issues. The beta version is updated one week before official app releases, to discover final issues. 
Feedback regarding these versions cannot be provided in form of app reviews. Instead, testers email their feedback directly to the development teams as indicated during sign up for the programs. This approach aims to decouple  testing (i.e., identifying and reporting bugs) from actually using and assessing the app. 

Further, software practitioners should select individual users to participate in A/B-tests (e.g. as both Duolingo and Spotify do, see https://bit.ly/2FA7U0o, https://bit.ly/2FHntT3, and https://bit.ly/2VYpf8h). One group of the users temporarily receives access to new or modified app features. Duolingo states: \textit{``Every week, we test at least 10 things on a portion of our users.''} (https://bit.ly/2HiB6KN).

Last, whenever possible new features should be gradually rolled out to all users so that app vendors can assess the overall impact on the emotional trend and react to unforeseen issues, e.g., by deactivating the functionality until the next app update.

\vspace{10px}
\noindent\fbox{
  \parbox{.95\columnwidth}{
    \textbf{Recommendation \#3:}\vspace{1mm}\\
    Software practitioners should explore app stores' functionality to distribute alpha and beta versions.
    The Apple App Stores allows to distribute these versions using TestFlight via email invite or public links (https://apple.co/1kxr08D). 
    On Google Play, developers can similarly release their apps using the Play console (https://bit.ly/1gLkkv2). Google Play allows to advertise alpha and beta versions on the official app description page, visible to all users.
    After testing, changes should be gradually rolled out to assess their overall impact on users' emotions and to be able to react to unforeseen issues.
  }
}


\subsection{Explain Changes to Users}

Major changes, such as the increase of the minimum required system version or the removal of app features, should be announced and explained to users. Studies show that users do not pay too much attention to release notes~\cite{McIlroy2016}. Instead, software practitioners should engage in conversations with users~\cite{bailey2019examining} or directly explain the changes within the app itself, e.g., using tutorials and tooltips.

We observed that apps that did not follow this lesson were affected by the pattern of \textit{steady decrease}. For example, for the CNN app a user reported: \textit{``What happened to local news. I checked that every day [...] please bring it back''}. For Microsoft Outlook the sentiment decreased significantly when users were affected by incompatibilities with new and old iOS versions, such as \textit{``Update [...] broke the app on iOS 8. Went back to using the stock mail app on my iPhone. Uninstalled it.''}. 

\vspace{10px}
\noindent\fbox{
  \parbox{.95\columnwidth}{
    \textbf{Recommendation \#4:}\vspace{1mm}\\
    App changes should be transparent and understandable to users. Therefore, major changes software practitioners do not want to release in smaller parts should be explained to users, e.g., using app built-in tutorials. Further, users with legacy devices and system versions should be redirected to alternatives  (e.g., web version) using announcements before stopping support.
  }
}


\subsection{Capture Implicit Feedback to Support Decisions}

Software practitioners should capture implicit feedback to empirically determine whether experimental app features should be integrated. At the beginning, the taken measures should reflect a basic overall goal (e.g., maximize the number of tracks listened for music apps, see https://bit.ly/2Mka18R). Then, more complex measures can be developed.

For example, Spotify performs A/B-tests even for unfinished features to decide whether these should be further developed. For testing, changes are split into atomic parts. When changing, e.g., the navigation, one test looks into the UI while another test focuses on the content, i.e., order of menu items (https://bit.ly/2Wb0NQw).

While implicit measures help software practitioners evaluate and optimize features against comparable criteria, explicit feedback provides additional information why taken measures change \cite{Garcia-Gathright:2018:MME:3240323.3241622}. For explicit feedback where no ratings exist, the sentiment can be calculated to quickly assess users' opinion. Further, it offers a broader understanding of the impact of changes. Users of no longer supported devices are, e.g. able to express their opinion in explicit feedback.

\vspace{10px}
\noindent\fbox{
  \parbox{.95\columnwidth}{
    \textbf{Recommendation \#5:}\vspace{1mm}\\
    Software practitioners should take further steps towards data-driven requirements engineering~\cite{7325177} by integrating logging frameworks, such as AppSee or Google Analytics, into their apps. Beginning with easy measures that relate to the app's overall goal, software practitioners should develop more complex ones by testing the impact of changes in atomic parts. The implicit measures complement explicit user feedback to support decisions which experimental features to integrate into apps.
  }
}


\section*{Acknowledgment}

This research was partially supported by the European Union Horizon 2020 project OpenReq under grant agreement no. 732463.


\begin{IEEEbiography}
    [{\includegraphics[width=1in,height=1.25in,clip,keepaspectratio]{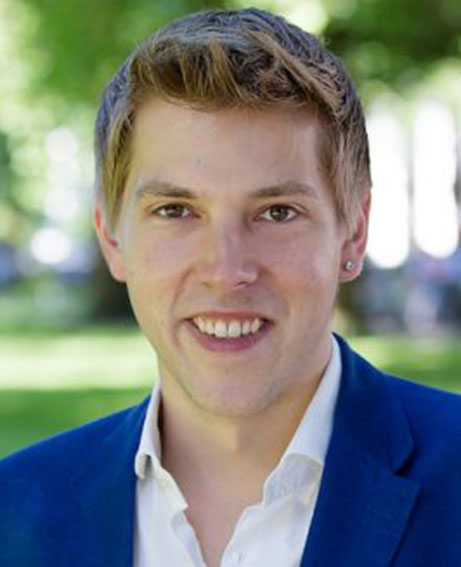}}]{Daniel Martens}
received the B.Sc. and M.Sc. degree in Computer Science from the University of Hamburg. He is a Ph.D. candidate in the Applied Software Technology group at the University of Hamburg. His research interests include user feedback, data-driven software engineering, context-aware adaptive systems, crowd-sourcing, and mobile computing. Besides his academic career, Daniel Martens also worked as a software engineer where he developed more than 30 top-rated iOS applications. He is a student member of the IEEE, ACM, and German Computer Science Society (GI). The photo of Daniel Martens was taken by UHH/Sukhina.
\end{IEEEbiography}

\begin{IEEEbiography}
    [{\includegraphics[width=1in,height=1.25in,clip,keepaspectratio]{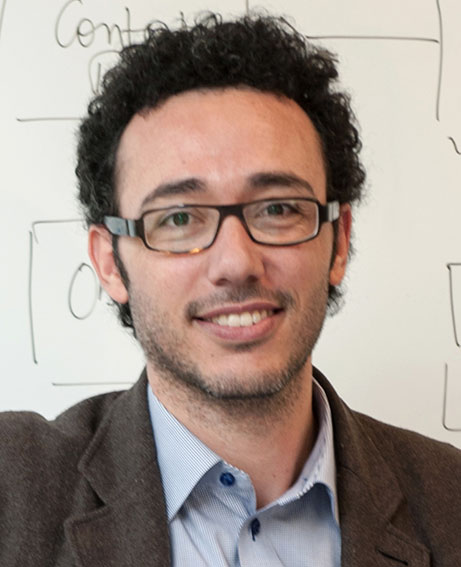}}]{Walid Maalej}
is a professor of informatics at the University of Hamburg where he leads the Applied Software Technology group. His research interests include user feedback, data-driven software engineering, context-aware adaptive systems, e-participation and crowd-sourcing, and software engineering’s impact on society. He received his Ph.D. in software engineering from the Technical University of Munich. He is currently a steering committee member of the IEEE Requirements Engineering conference and a Junior Fellow of the German Computer Science Society (GI). The photo of Prof. Walid Maalej was taken by Koepcke-Fotografie.
\end{IEEEbiography}

\bibliographystyle{IEEEtran}
\bibliography{refs}

\end{document}